\documentclass[aps,prl,twocolumn,groupedaddress]{revtex4}
\usepackage[top=1.9cm,left=1.8cm,right=1.8cm,bottom=1.9cm]{geometry}

\usepackage{amsmath}
\usepackage{amssymb}
\usepackage{color}
\usepackage{subfigure}
\usepackage{graphicx}
\usepackage{epstopdf}
\usepackage{dcolumn} 
\usepackage{comment}
\usepackage{ulem}
\usepackage{wasysym}

\DeclareMathAlphabet{\mathpzc}{OT1}{pzc}{m}{it}

\definecolor{darkgreen}{rgb}{0.25,0.75,0.25}
\definecolor{lightgray}{rgb}{0.75,0.75,0.75}

\newcommand{\cukD}{\hat{c}_{k,\uparrow}^\dag}
\newcommand{\cdkD}{\hat{c}_{k,\downarrow}^\dag}

\newcommand{\cuq}{\hat{c}_{k',\uparrow}}
\newcommand{\cdq}{\hat{c}_{k',\downarrow}}

\newcommand{\ket}[1]{\left| #1 \right>}

\newcommand{\ketBra}[1]{\left| #1 \right>\left< #1 \right|}

\newcommand{\expect}[1]{\left< #1 \right>}

\newcommand{\spectralfn}{\mathcal{S}}

\newcommand{\E}{\varepsilon}

\newcommand{\Z}{\hat{S}^{z}}

\newcommand{\Omup}{\Omega_{\uparrow}}
\newcommand{\Omdown}{\Omega_{\downarrow}}
\newcommand{\dup}{\delta_{\uparrow}}
\newcommand{\ddown}{\delta_{\downarrow}}
\newcommand{\dsig}{\delta_{s}}
\newcommand{\SigmaX}{\sigma^{x}}
\newcommand{\SigmaY}{\sigma^{y}}
\newcommand{\SigmaZ}{\sigma^{z}}
\newcommand{\SigmaV}{\hat{\vec{\sigma}}}
\newcommand{\Sveckk}{\hat{\vec{S}}_{{kk'}}}
\newcommand{\ex}[1]{\text{e}^{#1}}

\begin{document}

\title{Solid-State Spin-Photon Quantum Interface without Spin-Orbit Coupling}

\author{Martin Claassen}
\author{Hakan Tureci}
\author{Atac Imamoglu}
\affiliation{Institute for Quantum Electronics, ETH-Z\"urich, CH-8093 Z\"urich, Switzerland}

\begin{abstract}
We show that coherent optical manipulation of a single confined spin is possible even in the absence of spin-orbit coupling. To this end, we consider the non-Markovian dynamics of a single valence orbital hole spin that has optically induced spin exchange coupling to a low temperature partially polarized electron gas. We show that the fermionic nature of the reservoir induces a coherent component to the hole spin dynamics that does not generate entanglement with the reservoir modes. We analyze in detail the competition of this reservoir-assisted coherent contribution with dissipative components displaying markedly different behavior at different time scales and determine the fidelity of optically controlled spin rotations.
\end{abstract}

\maketitle

\vspace{2cm}

Quantum dot (QD) spins have emerged as a new paradigm for studying quantum optical phenomena in the solid-state. Motivated by potential applications in quantum information processing, the research in this field has focused on
understanding spin decoherence induced by the solid-state environment and implementing coherent spin manipulation \cite{cite:LossDiVincenco}. It has been proposed in this context that spin-orbit coupling provides a promising tool for implementing spin rotation by local electrical addressing of QD spins, mitigating the need for fast switching of local magnetic fields \cite{cite:coupledElectronSpins, cite:nowack}. Spin-orbit coupling plays an even more central role in interfacing spins with optical field. In fact, in all atomic as well as solid-state systems considered to date, optical manipulation of spins has been based on spin-orbit interaction either in the initial or final state of the optical transition \cite{cite:atac, cite:guptaAwschalom, cite:berezovskyAwschalom, cite:yamamoto, cite:SteelSham, cite:PazyZoller}.

In this work, we show that the non-Markovian dynamics of spin-exchange coupling between a confined spin and a
fermionic reservoir (FR) enables realization of a spin-photon interface \textit{in the absence of spin-orbit
interaction}. There are two key results conveyed by our work: first, we demonstrate that, contrary to the common
conception of treating reservoirs as sources of decoherence, an engineered fermionic reservoir gives rise to a
coherent contribution to single QD spin dynamics that could dominate over the decoherence it induces. Second, coherent optical manipulation of a single confined spin is possible in systems having weak or no spin orbit interaction. In addition to being interesting from a basic optical physics perspective, this latter result implies that confined spins in emerging material systems such as graphene where the spin-orbit effects are anticipated to be vanishingly small, could be manipulated using optical fields. With the exception of \cite{cite:Laird} where nuclear field gradients have been used for electrical manipulation of two-electron spin states, this is first discussion of electric field induced coherent spin rotation without spin-orbit interaction that we are aware of.

\begin{figure}[b]
	\centering
	\includegraphics[width=8.5cm]{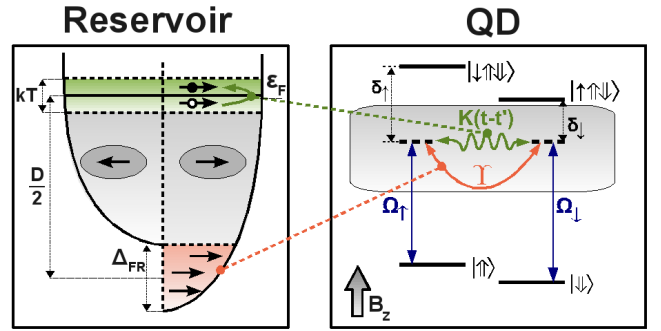}
	\caption{Reservoir-assisted spin-photon interface. QD: Lasers (Rabi freq. $\Omega_\downarrow$, $\Omega_\uparrow$) drive strongly detuned ($\delta_\uparrow$, $\delta_\downarrow$) spin-conserving dipole transitions ( $\ket{\Downarrow}\leftrightarrow\ket{\downarrow\Uparrow\Downarrow}$, $\ket{\Uparrow}\leftrightarrow\ket{\uparrow\Uparrow\Downarrow}$) between valence and excitonic spin states. Virtually excited states exhibit cotunnelling (dashed arrows) with a fermionic reservoir that induces both coherent ($\Upsilon$) and dissipative non-Markovian ($K(t-t')$) coupling of the two spin-transitions.
}
	\label{fig:scheme}
\end{figure}

We consider here the fidelity of FR-assisted Raman transitions of a generic qubit encoded in two hole spin states ($\ket{\Uparrow}, \ket{\Downarrow}$) of a valence-band orbital (of a QD, for example; see fig. \ref{fig:scheme}). Two lasers (with Rabi frequencies $\Omup$, $\Omdown$ and detunings $\dup$, $\ddown$) in Raman configuration couple the two hole spin states virtually to intermediate charged exciton (trion, $X^+$) states ($\ket{\uparrow\Uparrow\Downarrow}, \ket{\downarrow\Uparrow\Downarrow}$) involving two holes in a singlet state and a conduction band electron whose spin state is determined directly by the spin of the hole in the initial state \cite{footzb}. 
Charge fluctuations in the QD induced by  tunnel coupling to the FR are detrimental to spin coherence time and may be suppressed by moving to the local moment regime of parameters, in which the $X^+$ eigenstates lie sufficiently far below the Fermi edge such that it is energetically unfavorable for a QD electron to tunnel out or an extra electron to tunnel in. Employing a Schrieffer-Wolff transformation \cite{cite:schriefferwolff} to eliminate tunnel coupling to lowest order yields an effective Hamiltonian that couples the four eigenstates of the QD optical transitions with spin scattering excitations in the FR. In the limit of large laser detunings with respect to bare Rabi frequencies $\dup,\ddown \gg \Omup,\Omdown$, the relevant physics may be described by excitations of effective (dressed) hole spin states $\hat{\vec{\sigma}}$ to a virtual intermediate level exchange-coupled to FR. We obtain an effective spin exchange Hamiltonian $H_{sd} = \sum_{kk'} J_{kk'} \, \SigmaV \, \cdot \, \Sveckk$, a spin-conserving scattering term $H_{dir} = \sum_{kk'} \frac{W_{kk'}}{2} \left(\cukD\cuq+\cdkD\cdq\right)$ and an exchange interaction mediated energy shift of the effective hole spin states $H_0' = \sum_{kk'} \frac{J_{kk'}+W_{kk'}}{2}	$. Here $\Sveckk = \sum_{ss'} \hat{c}_{ks}^{\dagger}( \frac{1}{2} \vec{\tau}_{ss'}) \hat{c}_{k's'}$ ($\vec{\tau}$ are Pauli matrices, $\hat{c}_{ks}^{\dagger}$/$\hat{c}_{ks}$ fermionic creation/annihilation operators for the FR electrons with spin $s=\pm$). The optically induced exchange coupling to lowest order in $\Omega_{\sigma}(t)/\delta_{\sigma}$ reads
\begin{align}
    J_{kk'} &\approx \frac{\Omup(t)\Omdown(t) V^2}{4 \dup\ddown } \left[ \frac{1}{\Delta_d-\E_{k}} - \frac{1}{\Delta_0-\E_{k}} \right] + \mathbf{[ k \leftrightarrow k' ]}
\end{align}
and the corresponding directional coupling is $W_{kk'} = J_{kk'} + \Omup(t)\Omdown(t) V^2 / \left[2\dup\ddown (\Delta_d-\E_{k})\right] + \mathbf{[ k \leftrightarrow k' ]}$.
Here, we assume two-photon resonance condition between the hole spin states, namely $\dsig = \delta_{L} + s \Delta_e/2$ where $\Delta_e$ is the QD conduction band Zeeman splitting. $V$ is the tunnel coupling between QD and FR, $\Delta_d \approx \E_c+U_{ee}-2U_{eh} + \delta_L$ and $\Delta_0 \approx \E_c-2U_{eh} - \delta_L$ are the energies required to put an extra electron into or remove an electron from the QD. $U_{ee}$, $U_{hh}$ and  $U_{eh}$ are the intra-dot direct electron-electron, hole-hole and electron-hole Coulomb interaction energies in Hartree-Fock approximation and $\E_c$ is the bare QD conduction band electron energy. 

To implement coherent spin rotation, we assume that the FR spins are partially polarized along an axis that is tilted by an angle $\theta$ with respect to the QD spin axis ${\bf z}$. Such partial polarization can be achieved for example by tunnel coupling of the FR to ferromagnetic leads. The FR Hamiltonian is given by $H_{FR} =  \sum_{ks} \varepsilon_{k s} \hat{c}_{ks}^{\dagger}\hat{c}_{ks}$ with $\varepsilon_{ks} = \varepsilon_{k} + \frac{s}{2}\Delta_{FR}$, where $\Delta_{FR}$ is the FR Zeeman splitting, and spin $s$ denotes an electron spin along angle $\theta$. We introduce an effective finite bandwidth $D = 1/(2\rho)$ for FR electrons. Here, $\rho$ is the 2D density of states, and the band is symmetric around the Fermi edge $\E_F \equiv 0$. We assume weak spin polarization $\Delta_{FR} \ll D$. Within the second order Born approximation in $J_{kk'}$, we obtain a generalized master equation for the reduced dynamics of the QD hole spin. The corresponding Bloch equations are:
\begin{align}
    \frac{d}{dt}\SigmaZ &= -8 \int_0^t dt' K(t-t') \SigmaZ(t') + \sin(\theta)\Upsilon \SigmaX(t) \label{eq:BlochEqn1} \\
    \frac{d}{dt}\SigmaY &= -8 \int_0^t dt' K(t-t') \SigmaY(t') - \cos(\theta)\Upsilon\SigmaX(t) \label{eq:BlochEqn2} \\
    \frac{d}{dt}\SigmaX &= -8 \int_0^t dt' K(t-t') \SigmaX(t') \notag\\
    & \hspace{0.5cm}- \sin(\theta)\Upsilon\SigmaZ(t) + \cos(\theta)\Upsilon\SigmaY(t) \label{eq:BlochEqn3}
\end{align}
Here, $\sigma^{i}$ are the expectation values of the QD spin (axis ${\bf z}$) Pauli operators. Neglecting a spin-independent exchange interaction-mediated energy shift, the reduced QD spin dynamics for $\Delta_{FR}, T \ll D$ ($k_{B}=1$) described by Eqs.~\ref{eq:BlochEqn1}-\ref{eq:BlochEqn3} comprises a coherent spin precession quantified by the rate $\Upsilon = \sum_k \left(J_{kk} + W_{kk}\right) \expect{\Z_{kk}} $ and dissipative dynamics captured by the memory kernel $K(\tau) = {\rm Re} \sum_{kk'} \left|J_{kk'}\right|^2 \expect{ \Z_{k'k}(\tau)\Z_{kk'}(0)}$ \cite{footWeakPol}. $(\hat{\expect{O}} = {\rm Tr}\{\hat{O}~ \ex{-H_{FR}/T}\}/Z)$. In the following, we will discuss the case of orthogonal FR polarization ($\theta = \pi/2$).

Dissipative and coherent contributions are mediated by distinct parts of the FR. First, the action of the strip of width $\Delta_{FR}$ of excess spins at the lower edge of the FR may be viewed as generating a net magnetic moment of the 2DEG around which the QD spin precesses with rate
\begin{align}
    \Upsilon &\approx \frac{1}{2} \rho \left[ J_{-\frac{D}{2},-\frac{D}{2}} + W_{-\frac{D}{2},-\frac{D}{2}} \right] \cdot \Delta_{FR}
\end{align}
This action is mediated via energy-conserving co-tunnelling processes originating from this strip of electrons through an intermediate doubly-occupied QD. This process is coherent in the sense of keeping the QD spin states disentangled from the reservoir state continuum. Surprisingly, the lack of entanglement is not caused by the collective spin uncertainty of polarized FR electrons as seen from the QD. Rather, Pauli exclusion ensures that the absence of free spin states in the polarized strip prohibits modification of the reservoir. This coherent contribution to the system dynamics is valid even in the limit of a single polarized spin.

The dissipative part of the dynamics is mediated by electron-hole scattering processes within a strip of width $\sim  \text{max}(1/t \, , \, T)$ around the Fermi energy. This is, in essence, a dissipative limit of Kondo effect where the QD spin tries to evolve to a singlet with FR electrons but the coupling is too weak to maintain correlations, moving the system towards a maximum uncertainty state $\varrho = \left(\ketBra{\Uparrow} + \ketBra{\Downarrow}\right)/2$. Switching on the two lasers kicks the system into a non-equilibrium state accompanied by a shake-up of the Fermi Sea. The QD spin subsequently \textit{probes the full FR excitation spectrum} and moves in time through two distinctly different regimes of dissipative dynamics. To study this we consider the finite-bandwidth memory kernel $K(\tau) = \int_{-D}^D d\omega \cos(\omega\tau) \, \spectralfn(\omega)$, with FR spectral function $\spectralfn(\omega)$ given by
\begin{align}
    \spectralfn(\omega) = (J\rho)^2 \left[1+n_B(\omega)\right] T \log\left[\frac{\cosh\left(\frac{w-|w|+D}{4T}\right)} {\cosh\left(\frac{w+|w|-D}{4T}\right)}\right]
    \label{eq:SpectralFn}
\end{align}
in close connection with the spin-boson model \cite{cite:leggett, cite:rigorousBorn}. Here, $n_B(\omega) = 1/(e^{\omega/T}-1)$.
As implied by Kramers' theorem, there is no Lamb shift for degenerate dressed hole spin levels. We approximated $J_{kk'}$ by its value at the Fermi edge $J = J_{k_Fk_F}$
Microscopically, $\omega$ models the energy of electron-hole excitations in the reservoir.
A small-frequency expansion of spectral function (\ref{eq:SpectralFn}) gives $\spectralfn(\omega) = \frac{1}{4}(J\rho)^2 \left[ 2 kT + \omega + \mathcal{O}\left(\omega^2/T\right) \right]$. The corresponding memory kernel expansion is $K(\tau) = \gamma_M \cdot \delta(\tau) + (J\rho)^2/\tau^2$, with $\gamma_M = \frac{\pi}{2} (J\rho)^2 kT$ yielding the decay rate within Markov approximation, $\langle \hat{\sigma}^{i} (t) \rangle \sim \ex{-\gamma_M t}$. This exponential decay is the well-known Korringa relaxation \cite{cite:korringa} in the high-temperature limit. We find from the inspection of the expansion that exponential decay is only valid for $t \gg 1/T$.
What is not apparent within second order Born approximation is that the validity of this dynamical regime is bounded further by a time-scale $1/T_{K}$ where $T_{K}=D\sqrt{J\rho}e^{-1/2J\rho}$ is the Kondo temperature \cite{cite:TDNRG, cite:kondomodelheavyfermions} (for $T<T_{K}$, there is no time span within which the Markov approximation is valid). For $t \gg 1/T_{K}$ a FR screening cloud builds up over time and ultimately screens the localized spin into a singlet. Note that this indicates that the entire Born expansion fails - the distinction between system and reservoir is lost, a strongly correlated many-body state forms.

\begin{figure}[t]
    \centering
	\includegraphics[width=7cm,trim=1cm 0.5cm 1cm 1cm]{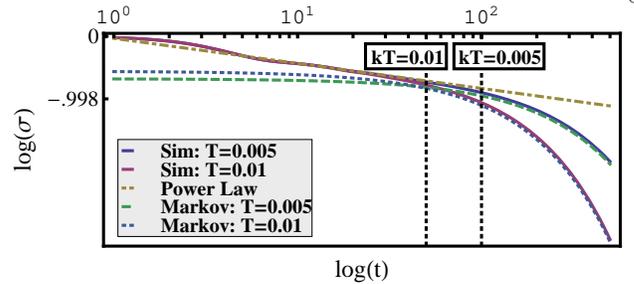}
	\caption{LogLog-Plot of $\SigmaZ(t)$, using $J\rho = 0.005$. Units scaled with respect to $D$. Solid lines show simulation results for $T=0.005D$ (blue) and $T=0.01D$ (red), dashed lines are intermediate- and long-time asymptotics. Vertical lines denote expected cross-overs at $1/2T$ between non-Markovian and Markovian regimes. $\SigmaZ(t)$ displays power-law relaxation for $t < 1/2T$, and long-time exponential decay for $t > 1/2T$.}
	\label{fig:plots}
\end{figure}

To obtain an analytic expression for $\sigma^i(t)$ in the (non-Markovian) intermediate time scale ($1/D \ll t \ll 1/T$), we set $\Upsilon=0$ and use a long-time asymptotic expansion of the zero-temperature limit of the memory kernel
\begin{align}
    K(\tau) = \frac{1}{8} (J\rho)^2 D^2 \cos\left(\frac{D\tau}{2}\right) {\rm sinc}^2\left(\frac{D\tau}{4}\right)
    \label{eq:LowTMemoryKernel}
\end{align}
(with ${\rm sinc}(x) = \sin(x)/x$). This form of $K(\tau)$ indicates initial-time oscillations on an ultra-short and experimentally irrelevant time scale of $1/D$. An analytic expression for $t \gg 1/D$  can be obtained by considering the analytic properties of the Laplace transform of the spin dynamics $\sigma(s) = s^{-1} \left[ 1 + 8 \mathcal{L}\{K(\tau)\}/s \right]^{-1}$ around $s=0$:
\begin{align}
    \sigma(s \rightarrow 0) & \approx \left(\frac{\tilde{J}}{J}\right)^2 \frac{1}{s} \cdot \frac{1}{1 - 4 (\tilde{J}\rho)^2 \cdot \log\left(\frac{4s}{D}\right)}
\label{eq:asymptoticSigmaLaplace}
\end{align}
where $\tilde{J} = J / \sqrt{1+4(J\rho)^2}$.
$\sigma(s\rightarrow 0)$ has two branch points at $s=\{0, \infty\}$ and we choose the branch cut along the negative real axis. The long time asymptotic behavior of $\sigma(t)$ is characterized by the functional form of $\sigma(s\rightarrow 0)$ along a keyhole contour around $s=0$ with a radius $\epsilon$ in the complex plane and can be obtained by the power-law expansion of the denominator using recursive relationship $\Lambda \log(s) = 1 - s^{-\Lambda} + \sum_{n=2}^\infty \frac{[-\Lambda \log(s)]^n}{n!}$ converging by choice of $\epsilon$ in powers of $\Lambda = 4\tilde{J}\rho$, which yields to lowest order in $\Lambda$ and in time domain:
\begin{align}
    \sigma(t\ll \frac{1}{T}) = \left(\frac{\tilde{J}}{J}\right)^2 \frac{1}{\Gamma\left(1-4 (\tilde{J}\rho)^2\right)} \left(\frac{D t}{4}\right)^{-4 (\tilde{J}\rho)^2}
\end{align}
with $\Gamma(x)$ the Gamma function. The prefactors account correctly for a slight dip in coherence during overdamped oscillations at times $<1/D$, greatly extending the short-time validity of the asymptotic calculations (see fig. \ref{fig:plots}).

The effect of initial powerlaw decay for $t\ll 1/T$ on the long-time dynamics $t
\gg 1/T$
may be extracted by approximation of $\spectralfn(\omega)$ by a piecewise linear function $\tilde{\spectralfn}(\omega < 2T) = T$ , $\tilde{\spectralfn}(\omega \geq 2T) = \lim_{T\rightarrow  0} \spectralfn(\omega)$, for which the long-time dynamics read
\begin{align}
	\sigma(t \gg \frac{1}{T}) = \left(\frac{\tilde{\gamma}_M}{\gamma_M}\right) e^{-8 \tilde{\gamma}_M t}
\end{align}
with modified relaxation rate $\tilde{\gamma}_M = \gamma_M / [1-4(J\rho)^2 \log(8T/D) ]$.

\begin{figure}[t]
    \centering
    \includegraphics[width=8.8cm,trim=1.7cm 2cm 0cm 2cm]{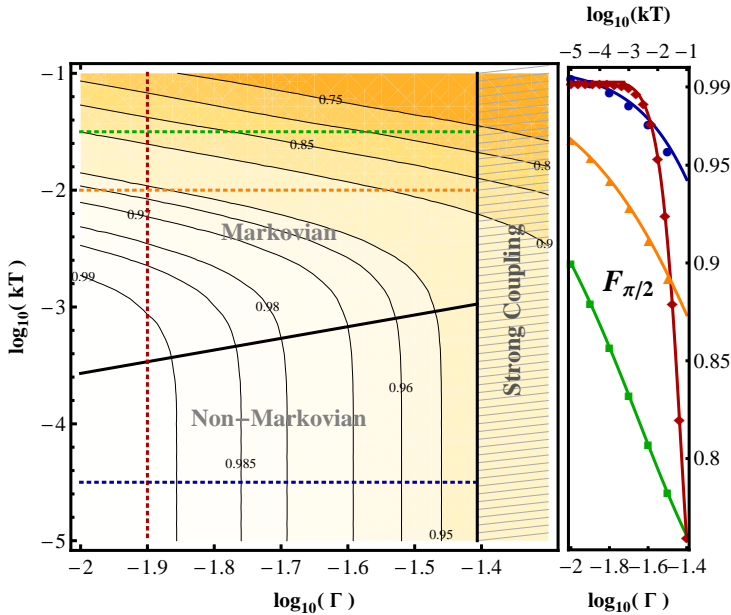}
    \caption{\textit{Left:} Contour plot of fidelity $F_{\pi/2}$ in Markovian and Non-Markovian regimes. Each pixel expresses the fidelity of the final state of a $\pi/2$ pulse after time $\tau_{\pi/2}(\Gamma) = \frac{\pi}{2\Upsilon(\Gamma)}$, given parameters $T$ (temperature) and  $\Gamma = \pi\rho\Omup\Omdown V^2/4\dup\ddown$. Symmetric model with $-\Delta_0 = \Delta_d = 0.5 D$ and $\Delta_{FR} = 0.1 D$ giving $J = 8\Gamma/\pi D$ and $\Upsilon = 8\Gamma\Delta_{FR}/3\pi D$. All energies are scaled by bandwidth $D$. Black lines display $\tau_{\pi/2} = 1/2T$ and $\tau_{\pi/2} = 1/T_K$ (vertical line). Simulation data is sampled with steps $0.25T$, $0.1\Gamma$. The cross-over to non-Markovian relaxation is clearly depicted by the 'bending down' of fidelity contour lines, conveying that low-temperature non-Markovian spin manipulation is solely limited by the strength of tunnel coupling $\Gamma$, manipulation of which displays competition of $\Gamma^2$-dependent power-law dephasing and $\Gamma$-dependent coherent spin rotation. \textit{Right: } Select contour plot cuts with fixed $\Gamma$ or $T$ (red,blue,green,orange dotted lines in both graphs) are compared to analytic results.} \label{fig:fidelity}
\end{figure}

We now discuss the quality of optical spin manipulation via Raman transition. Fidelity of spin manipulation is commonly characterized in terms of $\pi/n$ pulses (i.e. $\pi/n$ rotations of the QD spin). We focus on performing a $\pi/2$ rotation - more precisely, rotating an initialized spin $\ket{\Uparrow}$ to symmetric superposition $\frac{1}{\sqrt{2}}\left(\ket{\Uparrow}+\ket{\Downarrow}\right)$. A common measure of fidelity is $F_{\pi/2} = \sqrt{\frac{1}{2}\left(1+\expect{\SigmaX(\tau_{\pi/2})}\right)}$ \cite{cite:nielsenchuang} with $\tau_{\pi/2}$ the optimal time for a $\pi/2$-rotation. We state fidelity for rectangular pulse shapes and in the limit of separability of dissipative and coherent dynamics in the equations of motion (\ref{eq:BlochEqn1})-(\ref{eq:BlochEqn3}).
This decoupling is justified in the context of analyzing a $\pi/n$-pulse by noting that while the reservoir-mediated coherent drive dresses the QD effective hole spin states with splitting $\sim \Upsilon$, $\pi/n$-rotations probe only timescales $\lesssim 1/\Upsilon$ (and hence do not see the modification of $\spectralfn(\omega)$ for energies $\apprle \Upsilon$). This approximation breaks down for longer times. In a non-Markovian regime of operation $\pi/(2\Upsilon) \ll \min(1/T,1/T_K)$, the fidelity is given by:
\begin{align}
    F_{\pi/2} = \sqrt{\frac{1}{2} + \frac{\tilde{J}^2}{2 J^2 \Gamma(1-4(\tilde{J}\rho)^2) } \left(\frac{\pi D}{8\Upsilon}\right)^{-4 (\tilde{J}\rho)^2} }
    \label{eq:fidelityPowerlaw}
\end{align}
In the Markov regime $1/T \ll \pi/(2\Upsilon)$, it reads:
\begin{align}
	F^{\rm Markov}_{\pi/2} = \sqrt{\frac{1}{2} + \frac{\tilde{\gamma}_M}{2\gamma_M} e^{-4\pi\tilde{\gamma}_M / \Upsilon}  }
	\label{eq:fidelityExponential}
\end{align}
Figure \ref{fig:fidelity} plots the fidelity of a $\pi/2$-pulse across different parameter regimes. We emphasize that while the Markovian result would hint at vanishing dissipation with decreasing temperature, the system in fact approaches a fundamental limit of dissipation with power-law decay. In this sense, while operation in a low-temperature regime is beneficial, the non-Markovian nature of the reservoir in fact impairs the quality of spin manipulation.

Our results show that high-fidelity coherent spin manipulation is possible by coupling the QD spin to a polarized FR, provided that $\Upsilon(\Gamma,\Delta_{FR}) \gg T$. This description remains valid as long as $\Gamma$ is sufficiently reduced to ensure that the spin rotation is completed on a time scale shorter than the cross-over to strong coupling ($1/T_K$). On the other hand, since the reservoir-mediated optical spin-flip rate is proportional to $\Gamma$, it is desirable to increase the exchange coupling in such a way to satisfy $t^{-1} < T \sim T_K$ and thereby reduce the timescale of coherent spin-rotation. This suggests that as $T,T_K \rightarrow 0$, we can make sure that any finite FR spin polarization $\Delta_{FR}$ would be sufficient for coherent rotation.

We have presented a scheme for coherent optical manipulation of a single confined spin mediated through tunnel coupling to an electron gas. We find that {\sl reservoir engineering} of such an electron gas, achieved in our case by weak spin polarization, can modify the nature of system reservoir coupling drastically, exploiting the fermionic nature of the reservoir and turning it into a resource that allows for coherent spin manipulation. Pauli exclusion for the polarized spin states ensures that coherent spin exchange processes cannot write off information on the reservoir. We show that the microscopic description of such a reservoir at low temperatures gives rise to non-Markovian dynamics offering an attractive regime of attaining high-fidelity spin manipulation. On a fundamental level, our findings demonstrate that contrary to the common wisdom, \textit{spin-orbit interaction is not necessary} for realizing a spin-photon interface.

\begin{acknowledgments}
We thank J.Taylor for helpful discussions. This work was supported by Swiss NSF under Grant No. 200021-121757. HET acknowledges support from the Swiss NSF under Grant No. PP00P2-123519/1. AI acknowledges support from an ERC Advanced Investigator Grant.
\end{acknowledgments}



\begin{thebibliography}{9}
\bibitem{footzb}
We consider here a generic situation where valence band electrons can be characterized by a spin-like quantum number. In the case of Zinc-blende semiconductor QDs this is possible through the combined action of spin-orbit interaction and strain which results in heavy-hole band with a pseudo-spin $3/2$. In this case, the polarization of the incoming light additionally provides spin-selective transitions.
\bibitem{footWeakPol}
We note that above expression for the Bloch equations breaks for the unphysical case of $\Delta_{FR} \approx D$, in which dephasing processes differ significantly in $x,y,z$ basis.
\end{thebibliography}
\end{document}